%% file: main_arxiv.tex
\documentclass{article}[11pt]

\usepackage{amsmath}
\usepackage{amssymb}
\usepackage{amsthm}
\usepackage{wrapfig}
\usepackage{fullpage}
\usepackage{float}
\usepackage{graphicx}
\usepackage{latexsym}

\theoremstyle{definition}
\newtheorem{theorem}{Theorem}[section]
\newtheorem{lemma}[theorem]{Lemma}

\newcommand{\para}[1]{\vspace*{.1cm}\noindent\textbf{#1.}}

\def\Underline{\setbox0\hbox\bgroup\let\\\endUnderline}
\def\endUnderline{\vphantom{y}\egroup\smash{\underline{\box0}}\\}
\def\|{\verb|}

\begin{document}

\title{Hardness of Reconfiguring Robot Swarms with Uniform External Control in Limited Directions \thanks{This research was supported in part by National Science Foundation Grant CCF-1817602.}}

\author{David Caballero \and Angel A. Cantu \and Timothy Gomez \and Austin Luchsinger \and Robert Schweller \and Tim Wylie}

\date{}
%get rid of page number on the bottom
\clearpage\maketitle
\thispagestyle{empty}

\vspace*{-.5cm}
\begin{center}
Department of Computer Science\\University of Texas - Rio Grande Valley \\Edinburg, TX 78539-2999, USA\\
\{david.caballero01, angel.cantu01, timothy.gomez01, austin.luchsinger01, robert.schweller, timothy.wylie\}@utrgv.edu
\end{center}

\begin{abstract}
%We study a model where particles exist within a board and move single units based on uniform external forces. We investigate the complexity of deciding whether a single particle can be relocated to another position in the board, and whether a board configuration can be transformed into another configuration. We prove that the problems are NP-Complete with $1 \times 1$ particles even when only allowed to move in 2 or 3 directions.
Motivated by advances is nanoscale applications and simplistic robot agents, we look at problems based on using a global signal to move all agents when given a limited number of directional signals and immovable geometry.
We study a model where unit square particles move within a 2D grid based on uniform external forces.  Movement is based on a sequence of uniform commands which cause all particles to move 1 step in a specific direction.  The 2D grid board additionally contains ``blocked'' spaces which prevent particles from entry.  Within this model, we investigate the complexity of deciding 1) whether a target location on the board can be occupied (by any) particle (\emph{occupancy problem}), 2) whether a specific particle can be relocated to another specific position in the board (\emph{relocation problem}), and 3) whether a board configuration can be transformed into another configuration (\emph{reconfiguration problem}). We prove that while occupancy is solvable in polynomial time, the relocation and reconfiguration problems are both NP-Complete even when restricted to only 2 or 3 movement directions.  We further define a hierarchy of board geometries and show that this hardness holds for even very restricted classes of board geometry.
\end{abstract}

\input{intro.tex}
\input{preliminaries.tex}

\input{occupancy.tex}
\input{NPmembership2and3.tex}
\input{monotone.tex}
\input{reconfiguration.tex}
\input{conclusion.tex}

\nocite{Balanza-Martinez:2019:JCDCG3}
\bibliographystyle{amsplain}
\bibliography{limited}

\end{document}

%% file: intro.tex
\section{Introduction}
The tilt model, proposed by Becker et al. \cite{Becker:2014:AfSS}, has foundations in classical robot motion planning.  This model consists of a 2D grid of open and blocked spaces, called the ``board", along with a set of unit square pieces/tiles place at open board locations.  A sequence of ``tilts" push all the board pieces maximally in a specified cardinal direction.  A sequence of such tilts transforms the initial board configuration into a new configuration.  Some natural computational problems related to this model are those of \emph{occupancy}, \emph{relocation}, and \emph{reconfiguration}. \emph{Occupancy} is the problem of determining if there exists a sequence of tilts such that a specific, initially empty, board location may be occupied by a particle on the board.  \emph{Relocation} is the problem of whether a sequence of tilts exists to relocate a specific tile from location $a$ to location $b$.
\emph{Reconfiguration} asks if a sequence of tilts exists to transform board configuration $A$ to board configuration $B$ ( where each configuration specifies the location of all tiles on the board).   These problems were recently all shown to be PSPACE-Complete (in 4-directions) \cite{FullTiltSequel}.

% if a single polyomino larger than a $1\times 1$ is in the system \cite{Balanza:2019:SODA}.

Here, we discuss a variant of this model (introduced in \cite{Becker:2013:MUM}) in which particles only move 1 step per tilt in the specified direction, rather than maximally.  Figure \ref{fig:simple_example} shows a simple example.  We further consider these problems with limited usable directions (e.g. only tilting down and right), as well as considering the effect of limiting the complexity of the geometry of the open spaces of the board. For example, one limited type of geometry is that in which the open spaces form an ``$x/y$-monotone" shape.
%A specific class of boards used for our relocation problem is ``$x/y$-monotone'', which can also be called vertically/horizontally monotone.

\subsection{Previous Work}
The problems of Occupancy, Relocation, and Reconfiguration were first studied in \cite{Becker:2014:AfSS} in the full tilt model. In this work, the authors showed NP-hardness for the Occupancy Problem. Soon after, the authors of \cite{Becker:2014:ICRA} showed that finding the minimum move sequence for reconfiguring one configuration to another is PSPACE-Complete. Additional algorithmic, complexity, and logic work was done in \cite{Becker:2019:PCCAL}. All of these results used only $1 \times 1$ pieces. Later work in \cite{Balanza:2019:SODA} relaxed the constraint on tile size and showed the Relocation and Reconfiguration Problems were PSPACE-Complete when only a single $2 \times 2$ polyomino is allowed. Recent work strengthened these results and showed PSPACE-Completeness for all three problems even when only allowing $1 \times 1$ pieces \cite{FullTiltSequel}.  Additional work has focused on the application of the tilt model for the assembly of general shapes~\cite{Becker:2018:MICRO,Manzoor:2017:PSAP,Schmidt:2018:EPSA}, including \emph{universal constructors}~\cite{FullTiltSequel,Balanza:2019:SODA}, and  sorting polyominoes~\cite{Keldenich:2018:SORT}.

\subsection{Our Contributions}
We inverstigate some natural questions related to these problems and seek to  
find simple versions that are still computationally intractable. We remove the requirement for tiles to slide maximally and focus on unit movements as in \cite{Becker:2013:MUM}. In this model, the occupancy problem is solvable in polynomial time (Theorem~\ref{thm:easyOccupancy}), so we focus on the Relocation and Reconfiguration problems. We show intractability based on restricted  directions for both reconfiguration and relocation. % This limits the length of the move sequences to be polynomial so these problems are now in NP. Lastly we limit the complexity of board geometry. The hierarchy of board types we are exploring has already been established in previous work. 
We show that the Relocation problem is NP-Complete even when limited to  monotone board geometry~(Theorem~\ref{thm:monotone_relocation}) and the Reconfiguration Problem is NP-Complete when limited to Connected boards~(Theorem~\ref{thm:2dir_reconfig}).  A summary of our results are shown in Table~\ref{tab:results}.

\begin{table}[t]
  \centering
    \begin{tabular}{| c | c | c | c | c | c | c | c }
    \hline
  \textbf{Problem} & \textbf{Direct.} & \textbf{Geometry} & \textbf{Result} & \textbf{Theorem}\\
    \hline
    Occupancy & Any & All & P & Thm.~\ref{thm:easyOccupancy} \\
    \hline
    Relocation & 2, 3 & Monotone & NP-Complete & Thm.~\ref{thm:monotone_relocation}\\
    \hline
    Reconfiguration & 2 & Connected & NP-Complete & Thm.~\ref{thm:2dir_reconfig}\\
    \hline
  \end{tabular}
  \caption{An overview of results. The items in column \textbf{Problem} list the problems explored in this model. \textbf{Direct.} indicates how many directions of movement are allowed for the result. For the case of 2 directions we consider orthogonal directions. \textbf{Geometry} describe the type of board geometry used for the reduction (or allowed by the algorithm). The \textbf{Result} column shows the complexity of the problem with the theorem being found in column \textbf{Theorem}. }
  \label{tab:results}
\end{table}

%% file: preliminaries.tex
\section{Preliminaries} \label{Prelims}

%\para{Board, Configuration, and Step}
%Due to space constraints, we present informal definitions for the more straightforward terms used throughout this paper. See section~\ref{sec:formal_definitions} for more formal definitions of these terms.
%A \emph{board} is a rectangular region of the 2D square lattice which consists of \emph{open} and \emph{blocked} locations.
%A \emph{tile} is a labeled unit square that may exist on an open board location. Tiles with corresponding labels may attach to each other.
%A \emph{polyomino} is a collection of attached tiles. A $1\times1$ polyomino is called a tile.
%A \emph{configuration} is an arrangement of polyominoes on a board.
%A \emph{step} is a global signal which moves all polyominoes in a configuration one unit distance in a specified direction (North, East, South, or West) unless stopped by a blocked location. We say that a configuration can be \emph{directly reconfigured} into another via a step. Further, we say a configuration $C$ can be \emph{reconfigured} into $C'$ if there exists a sequence of direct reconfigurations beginning with $C$ and ending with $C'$ (this is denoted $C \rightarrow_* C'$).
%A \emph{step sequence} is a series of steps which can be inferred from a series of directions; each direction implying a step in that direction.

\para{Board}
A \emph{board} (or \emph{workspace}) is a rectangular region of the 2D square lattice in which specific locations are marked as \emph{blocked}.  Formally, an $m\times n$ board is a partition $B=(O,W)$ of $\{(x,y) | x\in \{1, 2, \dots, m\}, y\in \{1, 2, \dots, n\}\}$ where $O$ denotes a set of \emph{open} locations, and $W$ denotes a set of \emph{blocked} locations- referred to as ``concrete.'' We classify the different board geometries according to the following hierarchy:

\begin{itemize}\setlength\itemsep{-.2em}
    \item Connected: A board where the set of open spaces $O$  is a connected shape.
    \item Simple:  A connected board is  \emph{simple} if $O$ has genus-0.
    \item Monotone:  A simple board where $O$ is either horizontally or vertically monotone.
    \item Convex:  A monotone board where $O$ is both horizontally  and vertically monotone.
    \item Rectangular:  A convex board where $O$ is a rectangle.
\end{itemize}

%\para{Tiles}
%A tile is a labeled unit square centered on a non-blocked point on a given board.

 %Tiles are also referred to as polyominoes.  While we only consider polyominos of size $1\times 1$ in this paper, we use the more general term polyomino in our formal definitions to allow for the extension to larger pieces in future work.
%Formally, a tile is an ordered pair $(c,a)$ where \emph{c} is a coordinate on the board, and \emph{a} is an attachment label. Attachment labels specify which types of tiles will stick together when adjacent, and which have no affinity. For a given alphabet of labels $\Sigma$, and some \emph{affinity} function $G: \Sigma \times \Sigma \rightarrow \{0,1\}$ which specifies which pairs of labels attract ($G(a,b)=1$) and which do not ($G(a,b)=0$),  we say two adjacent tiles with labels $a$ and $b$ are \emph{bonded} if $G(a,b)=G(b,a) = 1$.

%\para{Polyomino}
%A \emph{polyomino} is a finite set of tiles $P = \{t_1, \ldots t_k\}$ that is 1) connected with respect to the coordinates of the tiles in the polyomino and 2) \emph{bonded} meaning the graph of tiles in $P$ with edges connecting bonded tiles is connected. A polyomino that consists of a single tile is informally referred to as a ``tile.''

\para{Configurations}
A configuration is an arrangement of tiles placed on open locations of a given board.  Formally, a \emph{tile} is a labeled unit square centered on a non-blocked point on a given board.  A \emph{configuration} $C=(B, P=\{P_1\ldots P_k\})$ consists of a board $B$ and a set of non-overlapping tiles $P$ that each do not overlap with the blocked locations of board $B$.

\para{Step} A \emph{step} is a way to turn one configuration into another by way of a global signal that moves all tiles in a configuration one unit in a direction $d \in \{N,E,S,W\}$ when possible without causing an overlap with a blocked position or another tile.  Formally, for a configuration $C=(B,P)$, consider the translation of all tiles in $P$ by 1 unit in direction $d$.  If no overlap with blocked board spaces occurs, then the new configuration is derived by first performing this translation.  On the other hand, for each tile for which the translation causes an overlap with a blocked space, temporarily add these tiles to the set of blocked spaces and repeat.  Once the translation induces no overlap with blocked spaces, execute the translation of the remaining non-blocked tiles to arrive at the new configuration.  If all tiles are eventually marked as blocked spaces, then the step transition does not change the initial configuration.  %If a configuration does not change under a step transition for direction $d$, we say the configuration is \emph{$d$-terminal}.  In the special case that a step causes a tile to leave the board, we  remove the tile from the configuration.

We say that a configuration $C$ can be \emph{directly reconfigured} into configuration $C'$ (denoted $C \rightarrow_1 C'$) if applying one step in some direction $d \in \{N,E,S,W\}$ to $C$ results in $C'$.  We define the relation $\rightarrow_*$ to be the transitive closure of $\rightarrow_1$ and say that $C$ can be \emph{reconfigured} into $C'$ if and only if $C \rightarrow_* C'$, i.e., $C$ may be reconfigured into $C'$ by way of a sequence of step transformations.  %A related concept that is the focus of previous work is the \emph{tilt} transformation in which a single direction $d$ tilt consists of the repeated application of a direction $d$-step until the configuration is $d$-terminal.  In this paper we focus on the step transition, but discuss connections to previous work using the tilt transformation.

\begin{figure}[tb]
  \begin{minipage}[t]{.24\columnwidth}
    \centering
    \includegraphics[width=.9\textwidth]{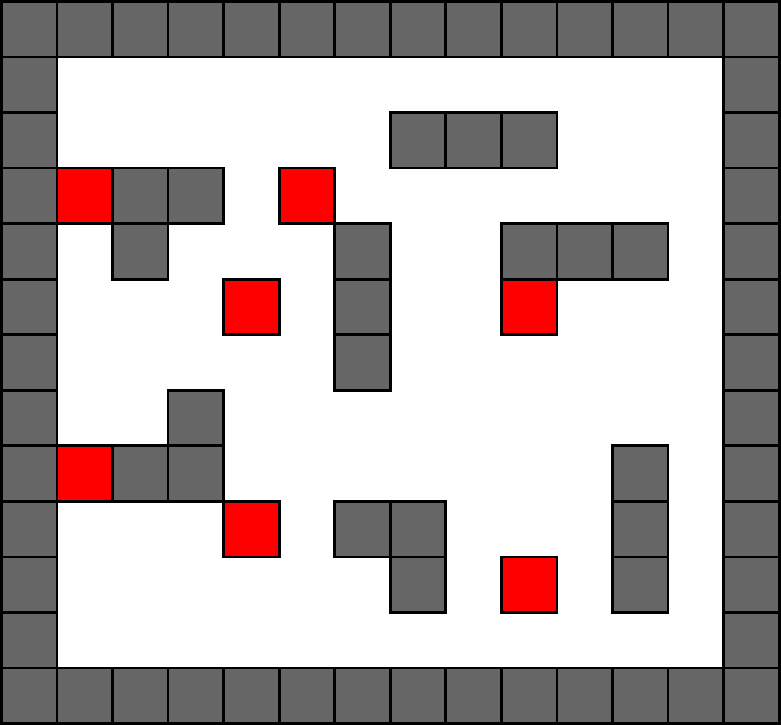}
    \text{Init}
  \end{minipage}
  \begin{minipage}[t]{.24\columnwidth}
    \centering
    \includegraphics[width=.9\textwidth]{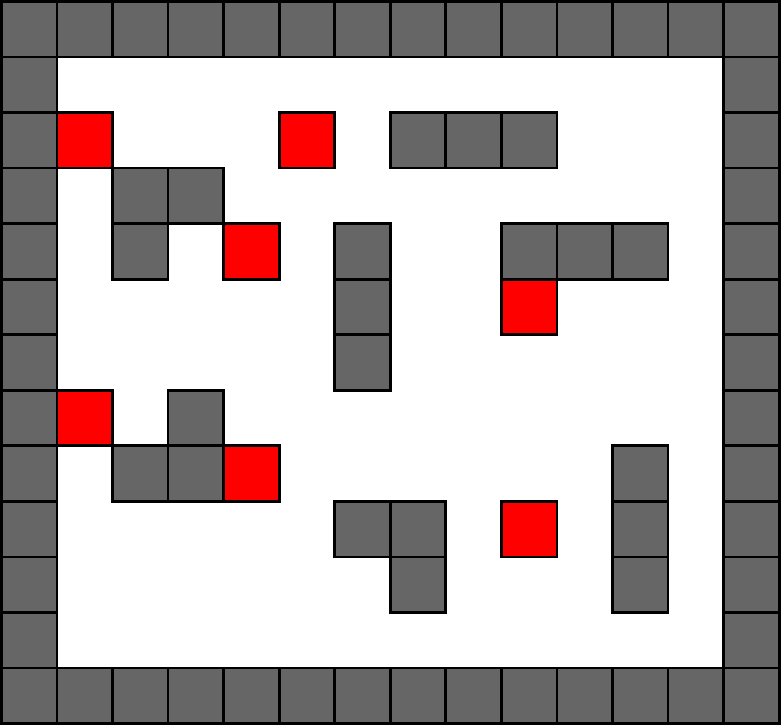}
    \text{$\langle N\rangle$}
  \end{minipage}
  \begin{minipage}[t]{.24\columnwidth}
    \centering
    \includegraphics[width=.9\textwidth]{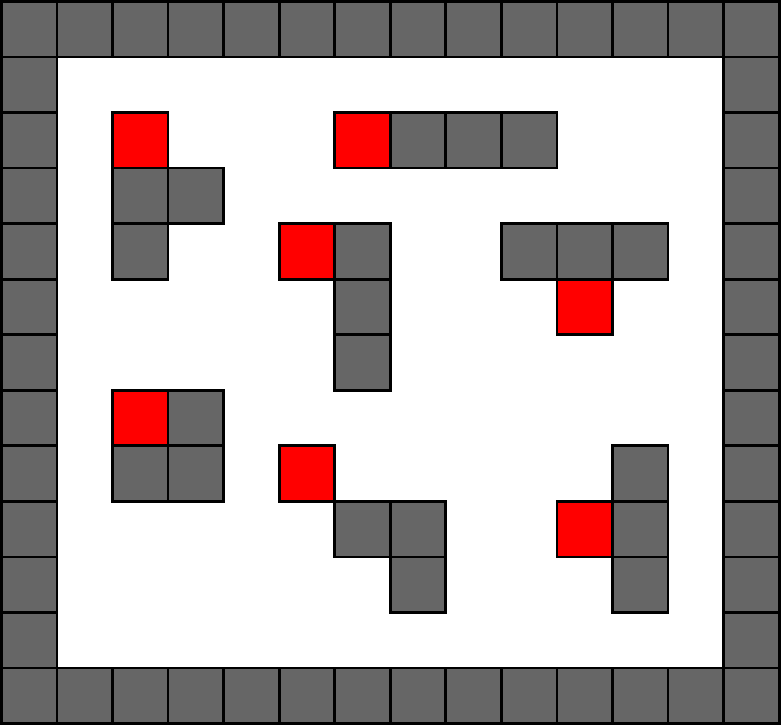}
    \text{$\langle E\rangle$}
  \end{minipage}
  \begin{minipage}[t]{.24\columnwidth}
    \centering
    \includegraphics[width=.9\textwidth]{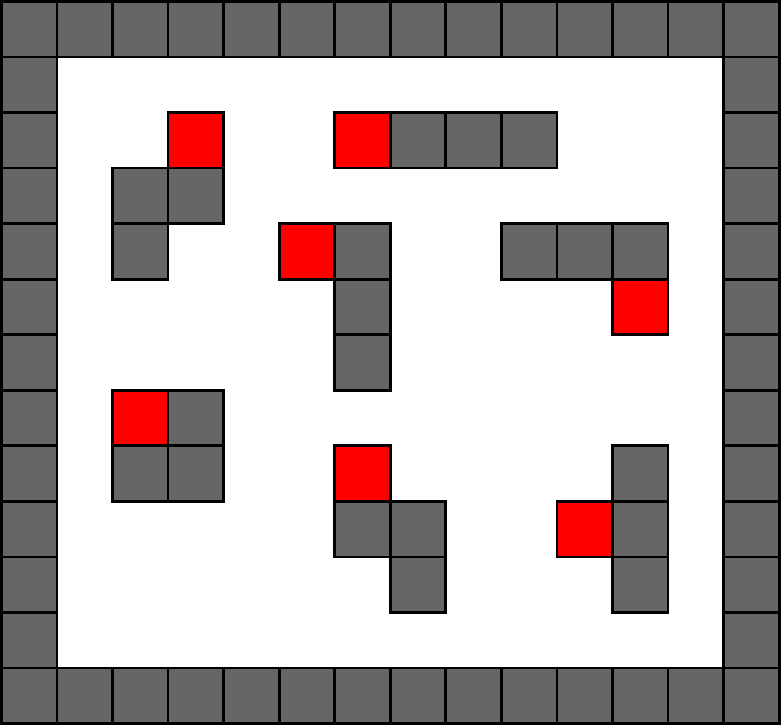}
    \text{$\langle E\rangle$}
  \end{minipage}
  \centering
  \caption{An example step sequence. The initial board configuration followed by the resulting configurations after an $N$ step, $E$ step, and then final $E$ step.}
  \label{fig:simple_example}
\end{figure}

\para{Step Sequence} A \emph{step sequence} is a sequence of steps which can be inferred from a sequence of directions $ D = \langle d_1, d_2,\dots, d_k \rangle$; each $d_i \in D$ implies a step in that direction. For simplicity,% when discussing a step sequence, 
we often just refer to the sequence of directions from which that sequence was derived. Given a starting configuration, a step sequence corresponds to a sequence of configurations based on the step transformation.  An example is shown in Fig. \ref{fig:simple_example}. %step sequence $\langle N, E, E \rangle$ and the corresponding sequence of configurations can be seen in Fig. \ref{fig:simple_example}.

\subsection{Problem Definitions}

In this paper we consider the following three problems under the step transformation.  In subsequent theorem statements we will state the complexity of these problems "under the step transformation" to help compare these results with work done "under the tilt transformation".

\textbf{Occupancy.} The occupancy problem asks whether or not a given location can be occupied by any tile on the board. Formally, given a configuration $C=(B,P)$ and a coordinate $e \in B$, does there exist a $C'$ such that $C \rightarrow_* C'$ where $C' = (B,P')$ and $\exists p \in P'$ that contains a tile with coordinate $e$?

\textbf{Relocation.} The relocation problem asks whether a specified tile can be relocated to a particular position. That is, given a configuration, a specific tile within that configuration, and a translation of that tile, does there exist a sequence of steps which moves the original tile to its translation?

\textbf{Reconfiguration.} The reconfiguration problem asks whether a configuration can be reconfigured into another. Formally, given two configurations $C = (B,P)$ and $C' = (B,P')$, does $C \rightarrow_* C'$?.

%% file: occupancy.tex
\section{Occupancy}

%\subsection{Occupancy with $1 \times 1$s}
\begin{theorem}\label{thm:easyOccupancy}
The occupancy problem under the step transformation is in P when using only $1\times 1$ tiles regardless of direction limitations.  In particular, the problem is solvable in $O(|B|)$ time, where $|B|$ denotes the number of positions on the input board $B$.
\end{theorem}

\begin{proof}
Perform a breadth-first-search from the goal location to determine if any position containing a tile is reachable by way of a sequence of unit distance steps between north/south or east/west connected open locations.  If no tiled position is reachable, the goal position is clearly not occupiable.  If there is a reachable tile, consider the closet such tile.  The shortest path connecting this tile to the goal location yields a tilt sequence that is guaranteed to place the tile at the goal, as the only way for it to be blocked would require another tile to move into position ahead of it along the shortest path, contradicting the claim that the first tile was the closest to the goal position.  Thus, the occupancy problem can be solved in $O(|B|)$ time. %answer is yes if and only if the goal location is connected to at least one tiled location, and can thus be solved in $O(|B|)$ time.
\end{proof}

%% file: NPmembership2and3.tex
\section{NP-Membership with Limited Directions}\label{sec:np}
In this section we prove NP membership for the relocation and reconfiguration problems when restricting steps to two orthogonal directions. The directions we consider, south and east, can be defined as the set $d = \{S,E\} $. Therefore, tiles can only move to a position that is south, east or southeast of the starting spot. We consider these directions since the case when the directions are opposite each other is solvable in polynomial time. For three directions we use west, south, and east, which can be represented by the set $d = \{W, S, E\} $. The following will consider a configuration with $n$ tiles and a board of size $l \times w$.

Membership comes from the polynomial upper-bound on the length of a move sequence using limited directions.

%\textbf{Maximal Configuration.} A configuration is Maximal for a set of directions if making a step in any of those directions results in the same configuration.

\begin{lemma}\label{lem:NP2}
For any two configurations $C$ and $C'$, if $C \rightarrow_* C'$ through a sequence of south and east steps, there exists a step sequence $S$ of length $\mathcal{O}(n (l + w))$ using only those two directions such that applying $S$ to $C$ results in $C'$.
\end{lemma}
\begin{proof}
Consider a tile $t$ at location $p$ in $C$ and at $p'$ in $C'$, the maximum distance between $p$ and $p'$ is $l + w$. Any step that moves $t$ brings it closer to $p'$ since $p'$ must be to the south east of it's starting location $p$, if a step moves $t$ further away from $p'$ than that step is not in $S$ since it would have moved $t$ to a location such that it can never reach $p'$. Now notice that each step must move at least one tile, and any step that moves a tile must move it closer to it's position in $C'$. Since each tile can move at most $l + w$ and each step must move at least one tile, if $C$ can be reconfigured to $C'$ then it will be able to do so in $\mathcal{O}(n (l + w))$ steps.
\end{proof}

\begin{lemma}\label{lem:NP3}
For any two configurations $C$ and $C'$, if $C$ can be reconfigured to $C'$ using only west, south, and east steps, there exists a step sequence $S$ of length $\mathcal{O}(nlw^2)$ using only those two directions such that applying $S$ to $C$ results in $C'$.
\end{lemma}
\begin{proof}
First let us consider the maximum number of south movements that can be made. Similar to above each south step moves at least tile further south so the maximum number of south steps that can be made is $nl$. Now considering the directions east and west, the maximum number of consecutive steps that can be made in one directions is $w$. Any step can be undone by making a step in the opposite direction unless the first step changed the position of tiles relative to each other. However this only brings the maximum move sequence for east and west to be $w^2$. Since the maximum move sequence before having to make a south step is $w^2$ and the maximum number of south steps is $nl$ the maximum move sequence length is  $\mathcal{O}(nlw^2)$.
\end{proof}

\begin{theorem}\label{thm:NP}
When restricted to 2 or 3 directions, the Relocation and Reconfiguration problems under the step transformation are in NP.
\end{theorem}
\begin{proof}
A step sequence can be used to verify positive answers to the Relocation and Reconfiguration problems. From Lemmas \ref{lem:NP2} and \ref{lem:NP3} we can see that the step sequences for the cases of 2 and 3 directions are polynomially bounded.
\end{proof}

%% file: monotone.tex
\section{Relocation with Limited Directions}
In this section we detail a reduction from 3SAT to relocation in monotone boards using limited directions.
We refer to the position the tile destined for relocation is initialized as location $a$ and the goal position as location $b$.
W.l.o.g, the directions used in the reduction are limited to $east$ and $south$.

\begin{figure}[tb]
  \begin{minipage}[t]{0.135\columnwidth}
    \centering
    \includegraphics[width=1.\textwidth]{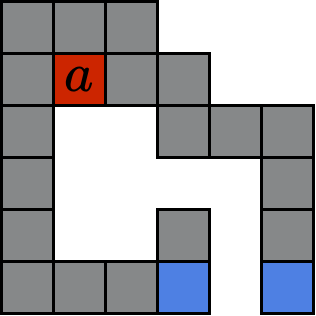}
    \text{(a)}
  \end{minipage}
  \begin{minipage}[t]{0.135\columnwidth}
    \centering
    \includegraphics[width=1.\textwidth]{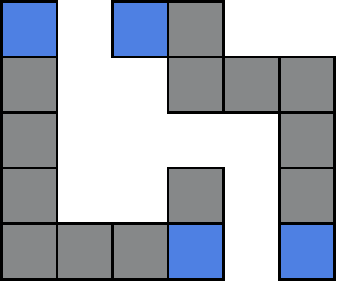}
    \text{(b)}
  \end{minipage}
  \begin{minipage}[t]{0.135\columnwidth}
    \centering
    \includegraphics[width=1.\textwidth]{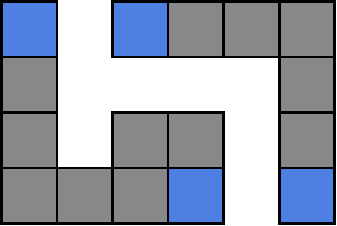}
    \text{(c)}
  \end{minipage}
  \begin{minipage}[t]{0.135\columnwidth}
    \centering
    \includegraphics[width=1.\textwidth]{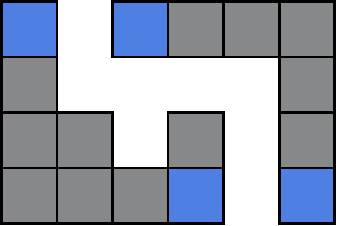}
    \text{(d)}
  \end{minipage}
  \begin{minipage}[t]{0.135\columnwidth}
    \centering
    \includegraphics[width=1.\textwidth]{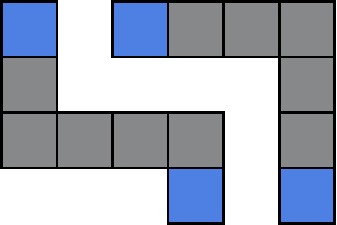}
    \text{(e)}
  \end{minipage}
  \begin{minipage}[t]{0.135\columnwidth}
    \centering
    \includegraphics[width=1.\textwidth]{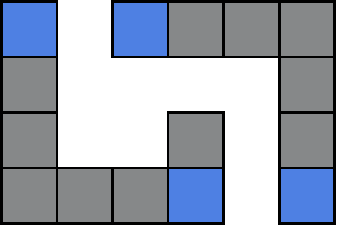}
    \text{(f)}
  \end{minipage}
  \begin{minipage}[t]{0.135\columnwidth}
    \centering
    \includegraphics[width=1.\textwidth]{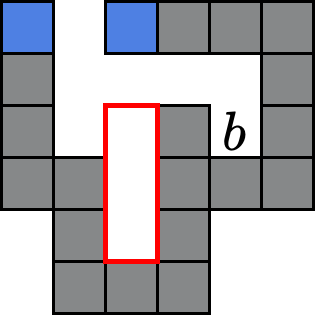}
    \text{(g)}
  \end{minipage}
  \caption{(a) Start gadget (b) Assignment gadget (c) Positive gadget (d) Negative gadget (e) Buffer gadget (f) Notch gadget (g) Goal gadget}
  \label{fig:gadgets}
\end{figure}

\subsection{Relocation Preliminaries}
We first describe the gadgets used for this reduction, as well as present some notation that we use throughout the section.

\paragraph{Gadgets}
All of the gadgets discussed in this section are show in Figure \ref{fig:gadgets}.
We will later describe how to construct a board by attaching gadgets together via their north (head) and south (foot) \emph{binding locations} (depicted in blue).
The relocation tile is located at position $a$ inside of the start gadget.
%Location $a$ of the board is allocated inside the Start gadget, including a tile occupying that location.
%A gadget is a structure of blocked spaces with two pairs of \emph{binding locations} on its north and south perimeter.
%Binding locations are blocked spaces two gadgets share when they are said to be bonded to one another.
%The north binding location pair is referred to as the \emph{head} and the south as the \emph{foot}.
%The Assignment gadget shares a similar architecture, where instead it includes a pair of north binding locations.
The Assignment gadget is where tilts will correspond to variable assignments for our reduction.
Every literal of a 3SAT formula has a corresponding Positive or Negative gadget, along with an associated tile that represents that literal.
%Assigning a literal to $true$ is achieved by moving its corresponding tile to the position in the Positive or Negative gadget which causes the tile to become confined.
%Assigning $false$ is done by evading the aforementioned position, leaving the tile unconfined.
Location $b$ resides inside the Goal gadget, a gadget whose structure grows with respect to the number of variables $N$.
The area within the Goal gadget that grows based on $N$ is outlined in Figure \ref{fig:gadgets}.
The Buffer and Notch gadget are utility gadgets discussed when needed. %used at specific moments during the reduction whose functionality will become apparent in the descriptions below.

\paragraph{Notation}
Here we present the notation for describing the construction of a tilt board with the gadgets described above.
We define $g_i \searrow g_j$ as gadget $g_i$'s foot binding with gadget $g_j$'s head.
A \emph{chain} is defined as a sequence of gadgets $S = \langle g_1, g_2, \ldots g_n \rangle$, s.t. $g_1 \searrow g_2$, $g_2 \searrow g_3$, $\dots, g_{n-1} \searrow g_{n}$.
It follows that a chain also has binding locations that allows it to bond with other chains or gadgets.
We define $G = \langle S_1, S_2, \ldots, S_n\rangle$ as a \emph{sequence of chains} s.t. $S_1 \searrow S_2$, $S_2 \searrow S_3$, $\dots, S_{n-1} \searrow S_{n}$.
Similarly, we say the binding of two sequences of chains $G = \langle S_1, S_2, \ldots, S_n \rangle$ and $G' = \langle S'_1, S'_2, \ldots, S'_m \rangle$ yields another sequence of chains $G'' = \langle S_1, S_2, \ldots, S'_{m-1}, S'_m \rangle$ s.t $S_1 \searrow S_2, \ldots, S_n \searrow S'_1, \dots, S'_{m-1} \searrow S'_{m}$.

\subsection{Board}
The board consists of three sections: the \emph{assignment}, \emph{formula}, and \emph{validation} section.
The tile initialized at location $a$ within the assignment section is referred to as the \emph{relocation} tile, the tiles inside the formula section are referred to as the \emph{literal} tiles, and the tiles within the validation section are called the \emph{validation} tiles. These board sections are all chains of the gadgets shown in Fig.~\ref{fig:gadgets}.

\paragraph{Validation Section}
The validation section is the sequence of gadgets $S_V = \langle g_1, g_2, \ldots, g_N\rangle$ where:

\begin{itemize}
\item $g_N = Goal$ $gadget$ and $\forall i < N$, $g_i$ = $Buffer$ $gadget$ with a validation tile allocated inbetween the head of every gadget.
\end{itemize}

\paragraph{Formula Section}
For the set of clauses $C$ and set of variables $X = \{x_1, x_2, \dots, x_N\}$, the clause chain of clause $c\in C$ is the sequence $G_c = \langle S_1, S_2, S_3, S_4, S_5\rangle$ where:

\begin{itemize}
\item $S_1 = \langle g_1, g_2, \ldots, g_N \rangle$ where if the literal $l_p\in c$ is the variable $x_p\in X$, then $g_p = \emph{Positive gadget}$ if $l_p$ is a positive literal or $g_p = \emph{Negative gadget}$ if $l_p$ is a negative literal.
The literal tile for literal $l_p$ is allocated inbetween the head of $g_1$.

\item $S_2 = \langle g_1, g_2, \ldots, g_N \rangle$ where if the literal $l_q\in c$ is the variable $x_q\in X$, then $g_q = \emph{Positive gadget}$ if $l_q$ is a positive literal or $g_q = \emph{Negative gadget}$ if $l_q$ is a negative literal.
The literal tile for literal $l_q$ is allocated inbetween the head of $g_1$.

\item $S_3 = \langle g_1, g_2, \ldots, g_N \rangle$ where if the literal $l_r\in c$ is the variable $x_r\in X$, then $g_r = \emph{Positive gadget}$ if $l_r$ is a positive literal or $g_r = \emph{Negative gadget}$ if $l_r$ is a negative literal.
The literal tile for literal $l_r$ is allocated inbetween the head of $g_1$.

\item $S_4 = \langle g_1, g_2, \ldots, g_{2N + 1}\rangle$ where $\forall i < {2N + 1}$, $g_i = \emph{Buffer gadget}$, and $g_{2N + 1} = \emph{Notch gadget}$.

\item $S_5 = \langle g_1, g_2, \ldots, g_{2N + 1}\rangle$ where $\forall i$, $g_i = \emph{Buffer gadget}$.\\
\end{itemize}

\begin{flushleft}
The formula section is therefore the sequence $G_F = \langle G_{c_1}, G_{c_2}, \ldots, G_{c_{|C|}} \rangle$ where every $G_{c_i} \in G_F$ is a clause chain and $G_{c_1} \searrow G_{c_2}$, $G_{c_2} \searrow G_{c_3}$, $\ldots$, $G_{c_{|C| - 1}} \searrow G_{c_{|C|}}$
\end{flushleft}

\paragraph{Assignment Section}
The assignment section is a sequence of chains $G_A = \langle S_1, S_2, S_3, S_4\rangle$ where:

\begin{itemize}
\item $S_1 = \langle g_1, g_2, \ldots, g_N\rangle$ where $g_1 = \emph{Start gadget}$ and $\forall i > 1$, $g_i = \emph{Assignment gadget}$.
The relocation tile is initialized within the Start gadget, as depicted in Figure \ref{fig:gadgets}.

\item $S_2 = \langle g_1, g_2, \ldots, g_{2N}\rangle$ where $\forall i$, $g_i = \emph{Notch gadget}$.

\item $S_3 = \langle g_1, g_2, \ldots, g_{2N + 1}\rangle$ where $g_1, g_{N+1}, g_{2N+1} = \emph{Assignment gadget}$ and $\forall i$ s.t $g_i \neq \emph{Assignment gadget}$, $g_i = \emph{Notch gadget}$.

\item $S_4 = \langle g_1, g_2, \ldots, g_{v}\rangle$ where $v$ = $|S_V| + |G_F|$, and $\forall i \leq v$, $g_i = \emph{Notch gadget}$.
\end{itemize}

\begin{flushleft}
The board $B = \langle G_A, G_F, S_V \rangle$ is therefore the combination of the three sections where $G_A \searrow G_F$ and $G_F \searrow S_V$.
\end{flushleft}

\subsection{Reduction}
The reduction can be understood as a two phase process.
In the first phase variables are given a truth value one-by-one in ascending order for the set of variables $X = \{x_1, x_2, \dots, x_N\}$.
The second phase consists of verifying if the variable assignments from the first phase satisfied the clauses of the formula.
Every clause chain is checked for any unconfined literal tiles, where if a clause chain confines at least one literal tile it is said to be satisfied, and unsatisfied otherwise.
A satisfied clause chain can have up to two remaining unconfined literal tiles, where an unsatisfied clause will have three unconfined literal tiles.
In order to identify satisfied and unsatisfied clauses, the remaining literal tiles are `counted' by attempting to occupy two open spaces using the literal tiles in each of their respective clause chains.
Attempting to occupy the two spaces with three literal tiles (e.g, an unsatisfied clause chain) will leave one literal tile unconfined which will be used to prevent relocation.

\begin{figure}[tb]
  \begin{minipage}[t]{.49\columnwidth}
    \centering
    \includegraphics[width=1.\textwidth]{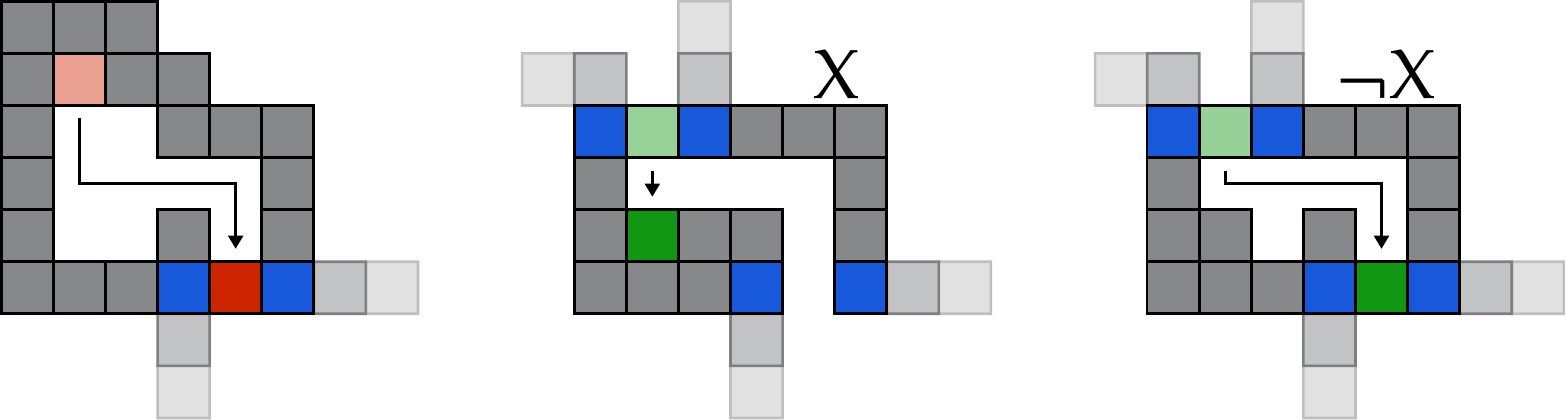}
    \text{(a) $\langle S^2, E^3, S^2\rangle$}
  \end{minipage}
  \begin{minipage}[t]{.49\columnwidth}
    \centering
    \includegraphics[width=1.\textwidth]{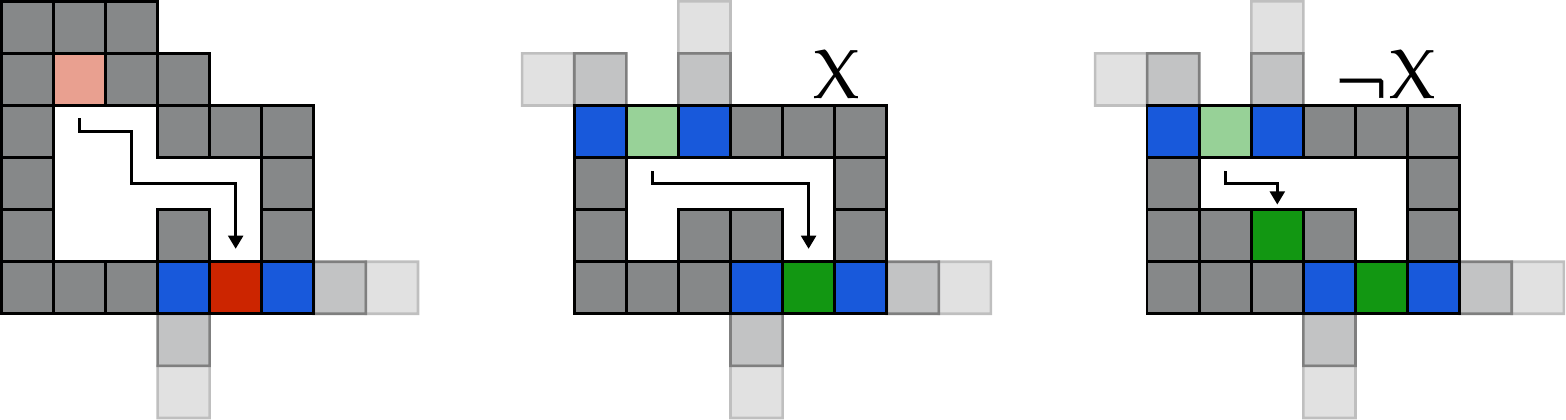}
    \text{(b) $\langle S, E, S, E^2, S^2 \rangle$}
  \end{minipage}
  \caption{Assigning (a) \emph{true} or (b) \emph{false} to some variable $x$. Setting a literal to true is done by confining its corresponding literal tile within a Positive or Negative gadget.}
  \label{fig:trueFalse}
\end{figure}

\paragraph{First Phase}
Starting with the board configuration described above, each variable in ascending order can be assigned some truth value by executing the step sequences depicted in Figure \ref{fig:trueFalse}.
During this phase, the relocation tile will visit every gadget inside the first chain of the assignment section if either one of the sequences depicted is executed for every variable truth assignment.

\begin{figure}[tb]
    \centering
  \begin{minipage}[t]{.32\columnwidth}
    \centering
    \includegraphics[width=1.\textwidth]{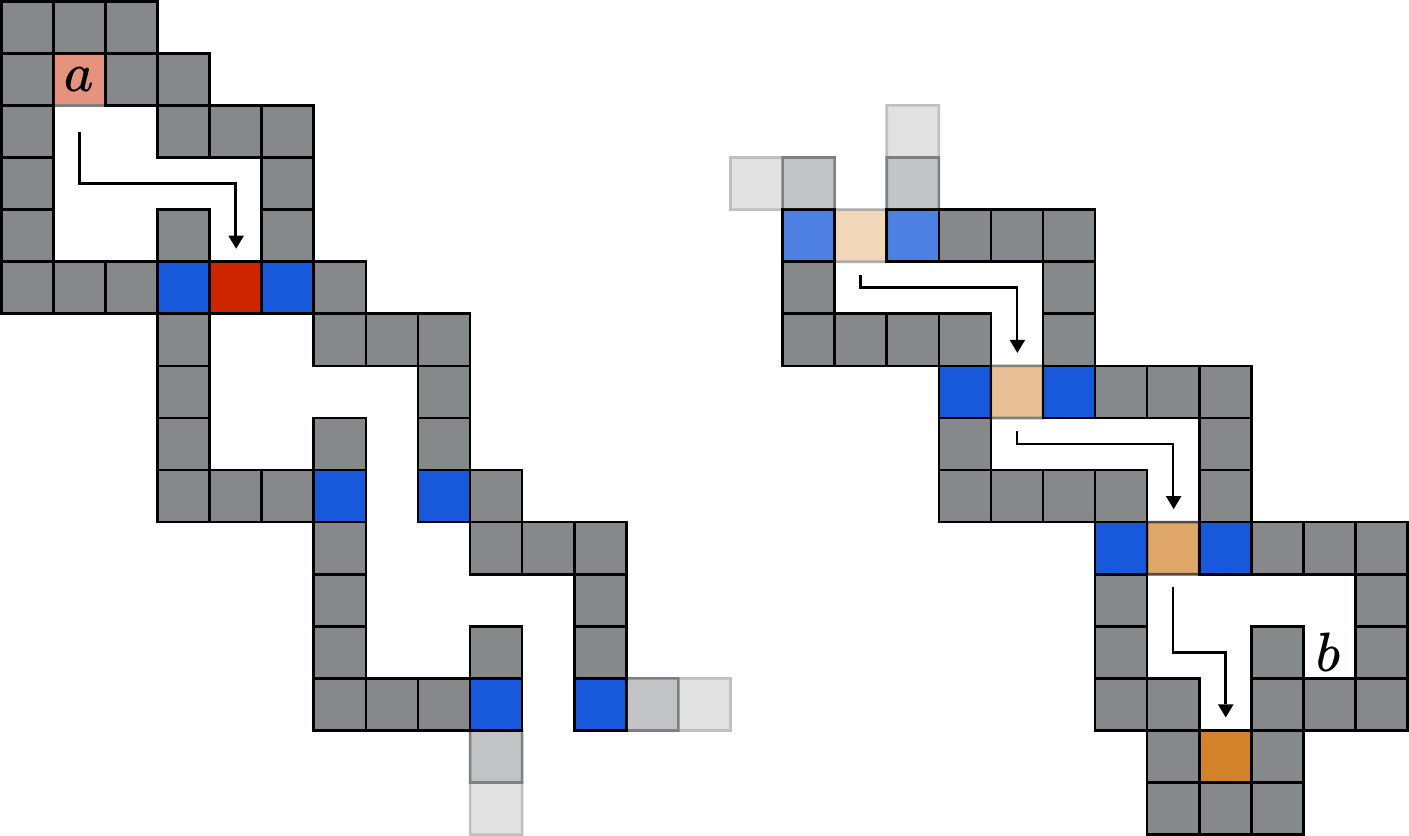}
    \text{(a)}
  \end{minipage}
  \begin{minipage}[t]{.32\columnwidth}
    \centering
    \includegraphics[width=1.\textwidth]{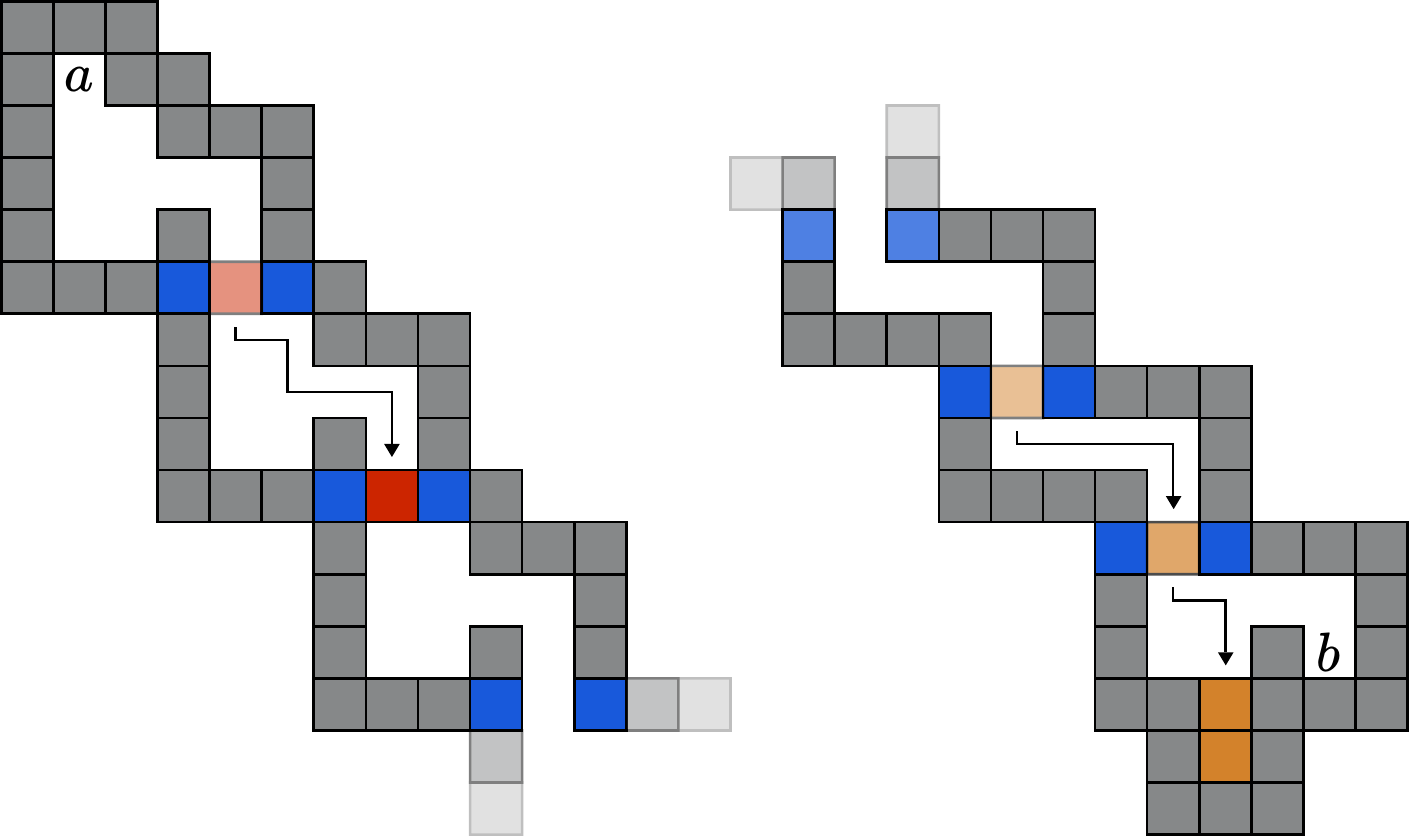}
    \text{(b)}
  \end{minipage}
  \begin{minipage}[t]{.32\columnwidth}
    \centering
    \includegraphics[width=1.\textwidth]{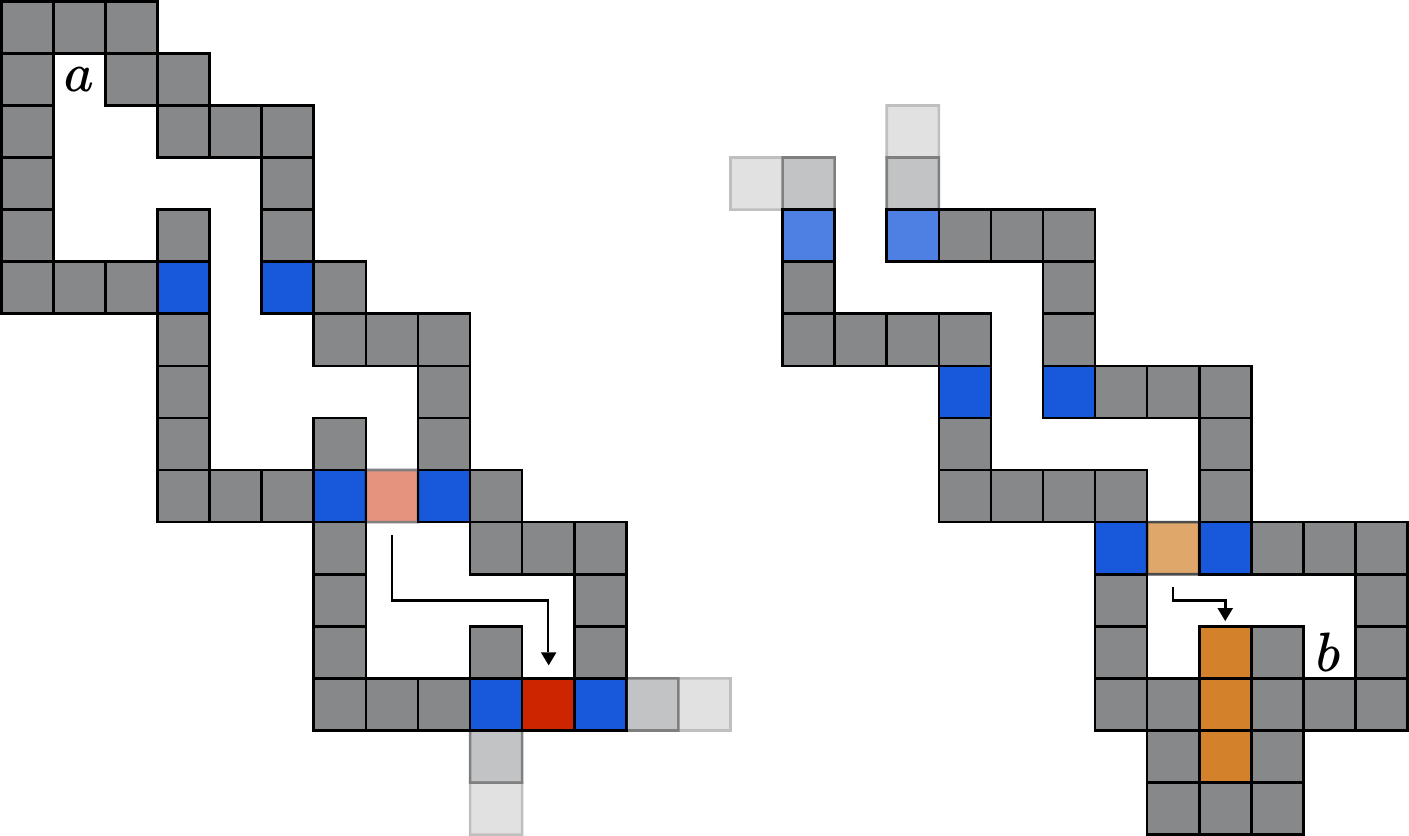}
    \text{(c)}
  \end{minipage}
  \caption{The leftmost chain belongs to the assignment section, and rightmost belongs to the validation Section. Validation tiles enter the Goal gadget for every assignment and Start gadget in the assignment section. This prevents any unnecessary step sequences from being performed.}
  \label{fig:validation}
\end{figure}

To ensure the step sequences depicted are the only ones executable during this phase, the validation tiles in the validation section will enter the Goal gadget one-by-one for every assignment in order to risk the unwanted occupancy of location $b$ with one of the validation tiles.
The validation tiles must solely experience the step sequences depicted above, unless the validation tiles will enter the notch hallway towards location $b$ inexorably, as depicted in Figure \ref{fig:validation}.
This relation between the relocation and validation tiles ensures that the two sequences depicted are the only ones performed during this phase.

\begin{figure}[]
  \begin{minipage}[t]{.32\columnwidth}
    \centering
    \includegraphics[width=1.\textwidth]{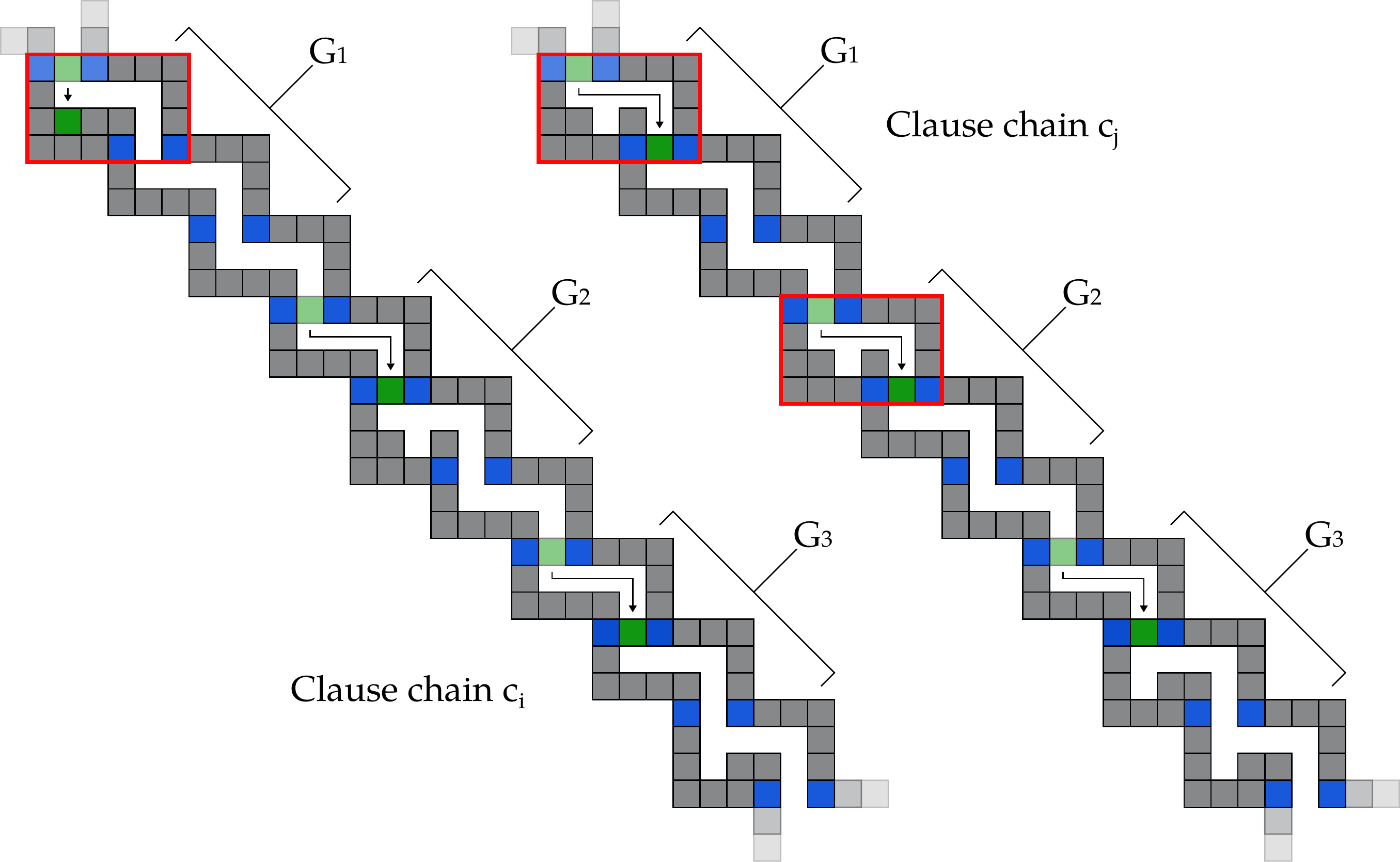}
    \text{(a) $x_1 \rightarrow 1$}
  \end{minipage}
  \begin{minipage}[t]{.32\columnwidth}
    \centering
    \includegraphics[width=1.\textwidth]{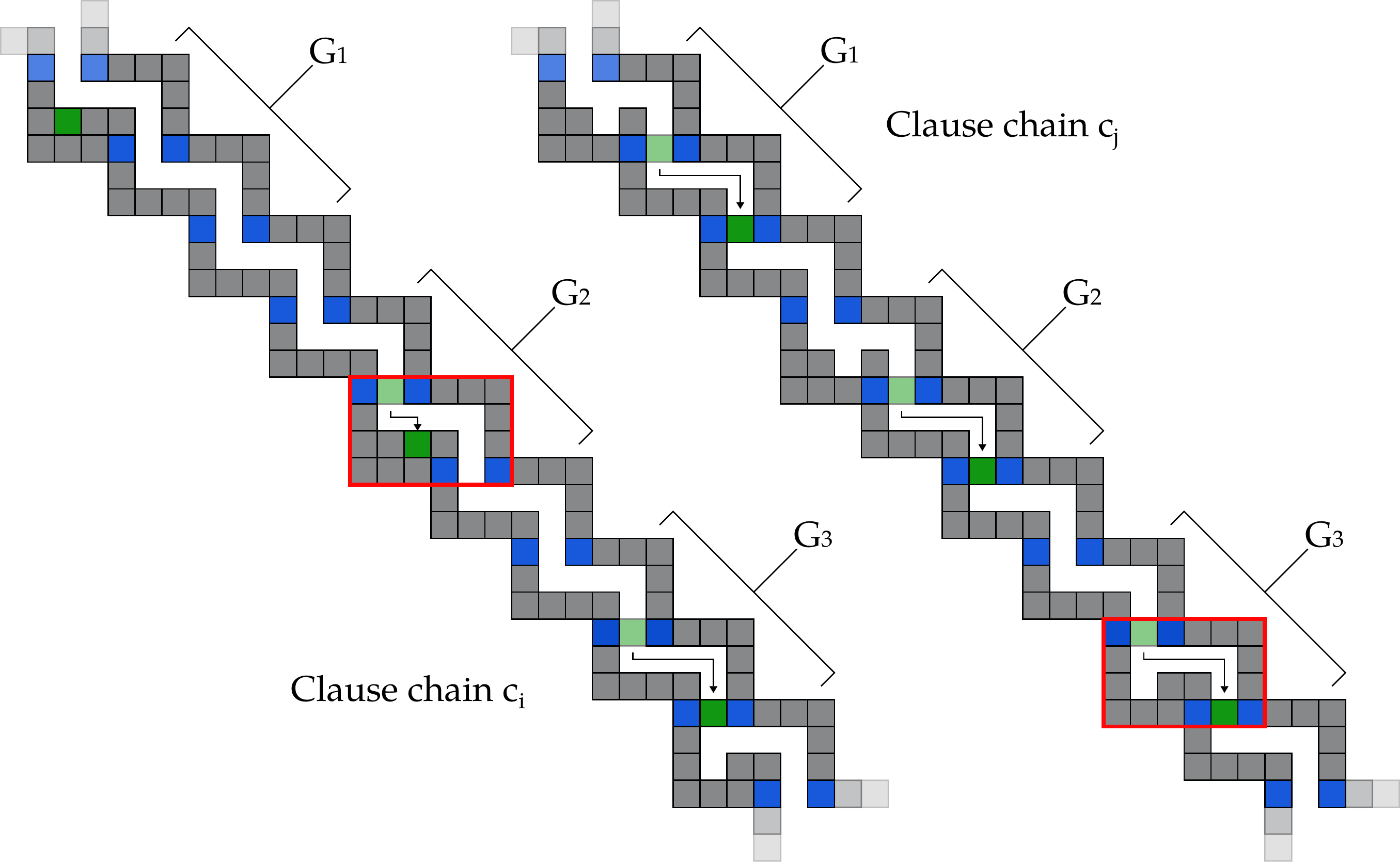}
    \text{(b) $x_2 \rightarrow 0$}
  \end{minipage}
  \begin{minipage}[t]{.32\columnwidth}
    \centering
    \includegraphics[width=1.\textwidth]{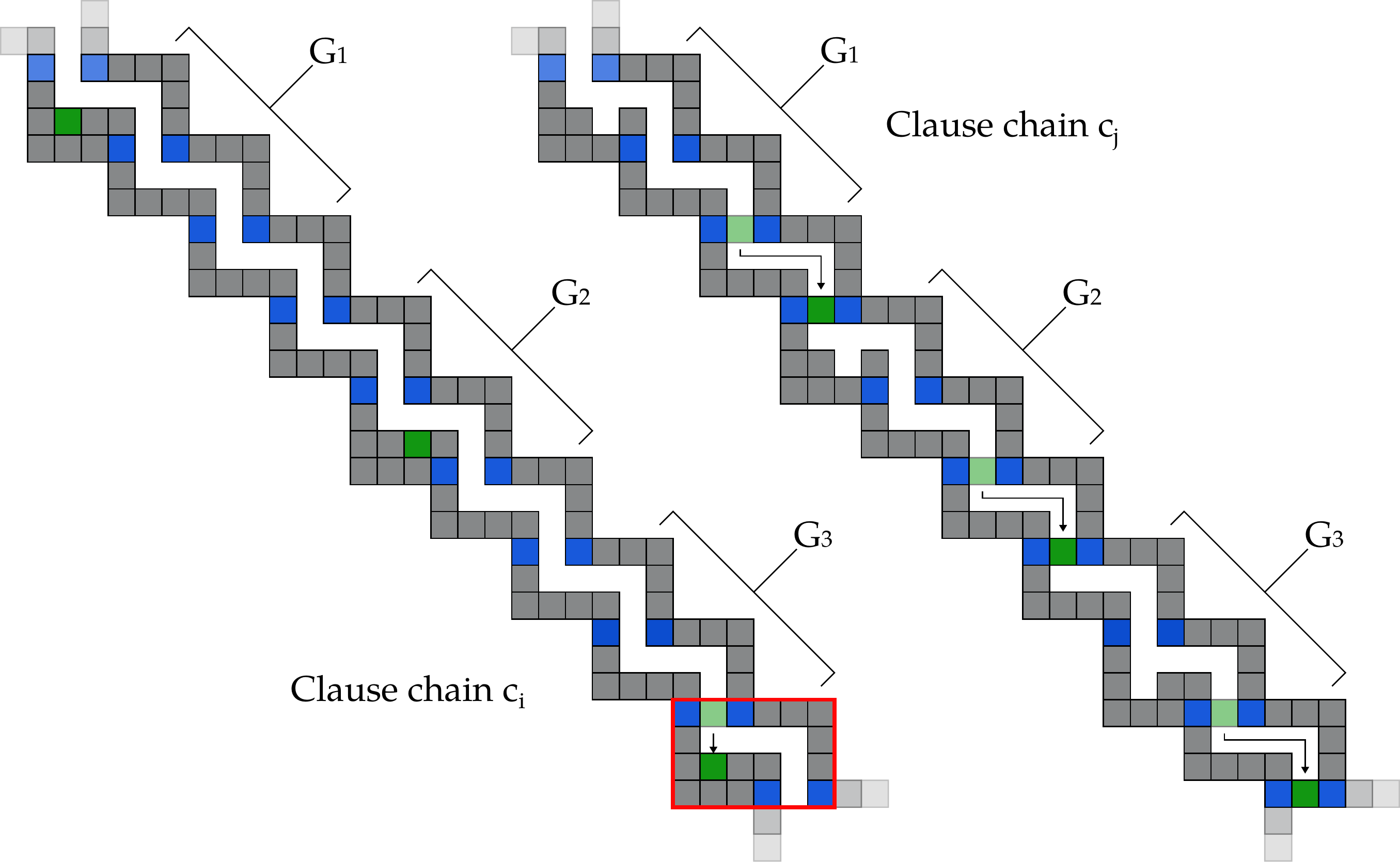}
    \text{(c) $x_3 \rightarrow 1$}
  \end{minipage}
  \centering
  \caption{Visualization of two clause chains $c_i = (x_1 \lor \neg x_2 \lor x_3)$, $c_j = (\neg x_1 \lor \neg x_1 \lor x_2)$ for $X = \{x_1, x_2, x_3\}$ during the first phase.}
  \label{fig:clauseChains}
\end{figure}

All literal tiles visit every gadget of the chain it resides in for every clause chain (one of which is a Positive or Negative gadget) as the step sequence depicted above is performed.
By design, the literal tiles reach a Positive or Negative gadget when its corresponding variable is next in the truth assignment procedure (Figure \ref{fig:clauseChains}).
After the last variable is assigned some truth value, every clause chain will confine at least one literal tile if the corresponding clause was satisfied, and zero literal tiles if it was not.

\paragraph{Second Phase}
Immediately after the first phase the relocation tile will reside inside the second chain of the assignment section, whose structure consists of only Notch gadgets.
The Notch gadget serves two functions in this phase, one of which is to forbid the movement in the $south$ direction when the relocation tile traverses through it.
The first step in this phase is therefore to execute the step sequence $\langle S, W^3, S^2 \rangle$ until the relocation tile reaches the third chain, which simultaneously moves any unconfined literal tiles to the fourth chain of their respective clause chains.

\begin{figure}[tb]
    \centering
  \begin{minipage}[t]{.32\columnwidth}
    \centering
    \includegraphics[width=1.\textwidth]{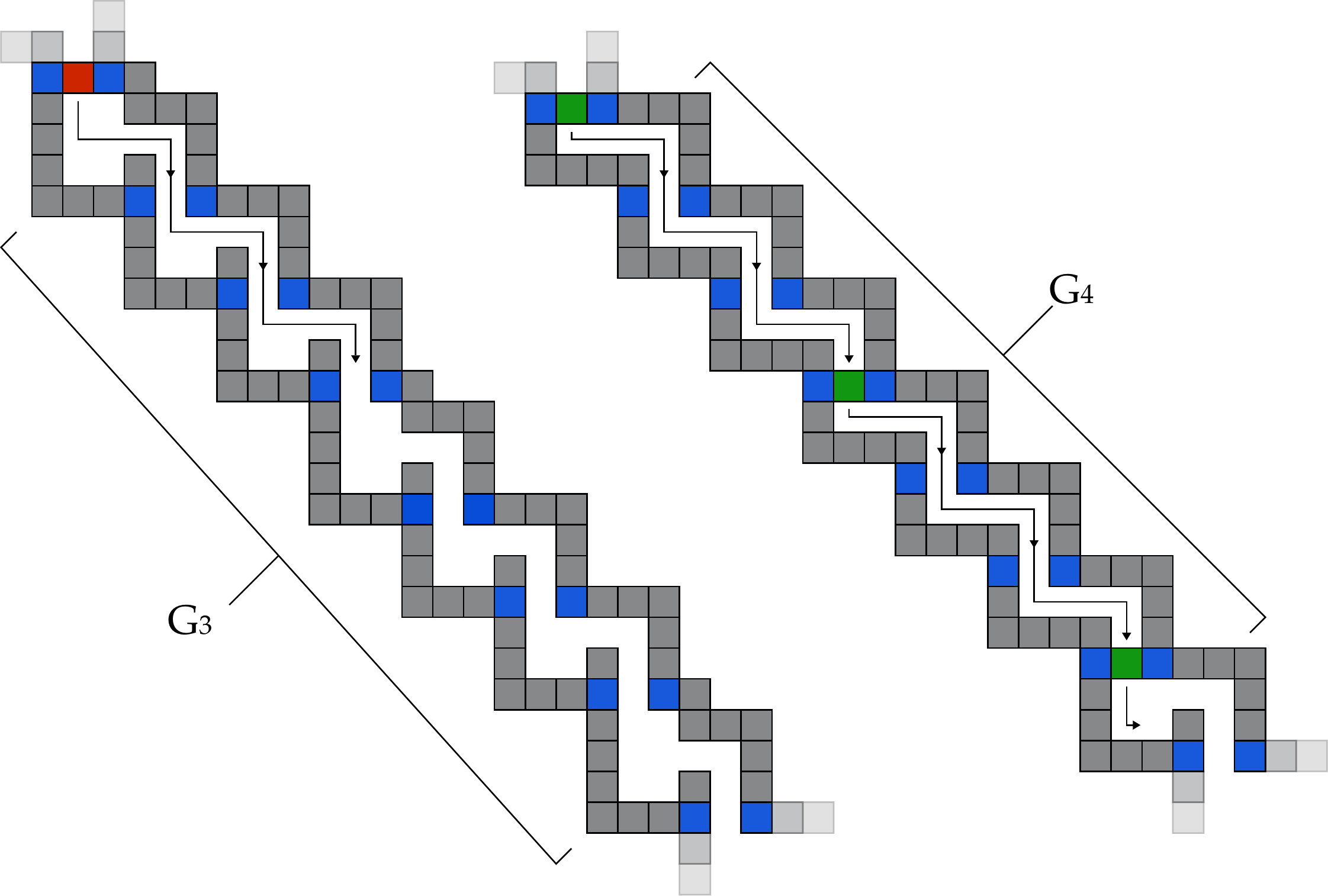}
    \text{(a)}
  \end{minipage}
  \begin{minipage}[t]{.32\columnwidth}
    \centering
    \includegraphics[width=1.\textwidth]{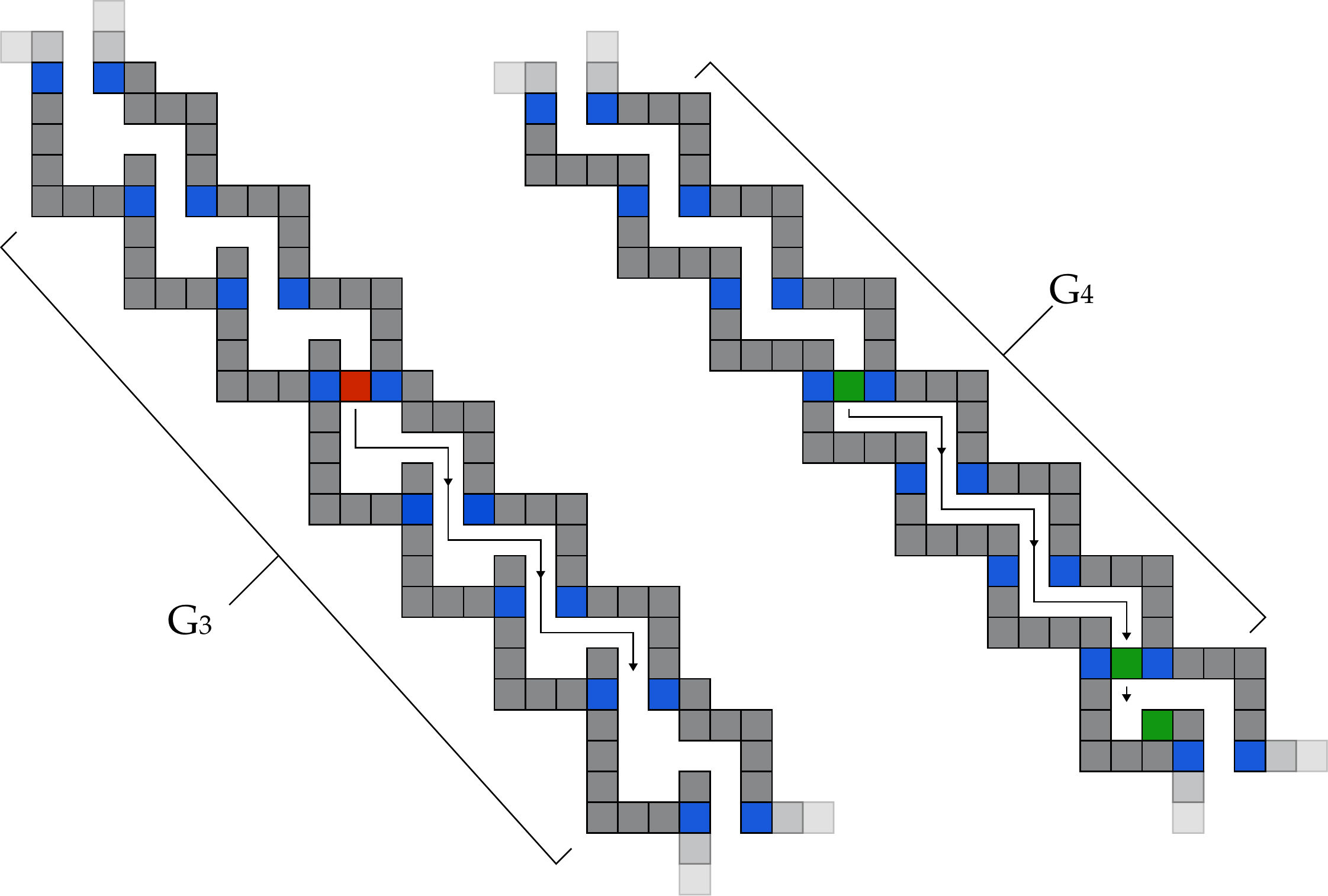}
    \text{(b)}
  \end{minipage}
  \begin{minipage}[t]{.32\columnwidth}
    \centering
    \includegraphics[width=1.\textwidth]{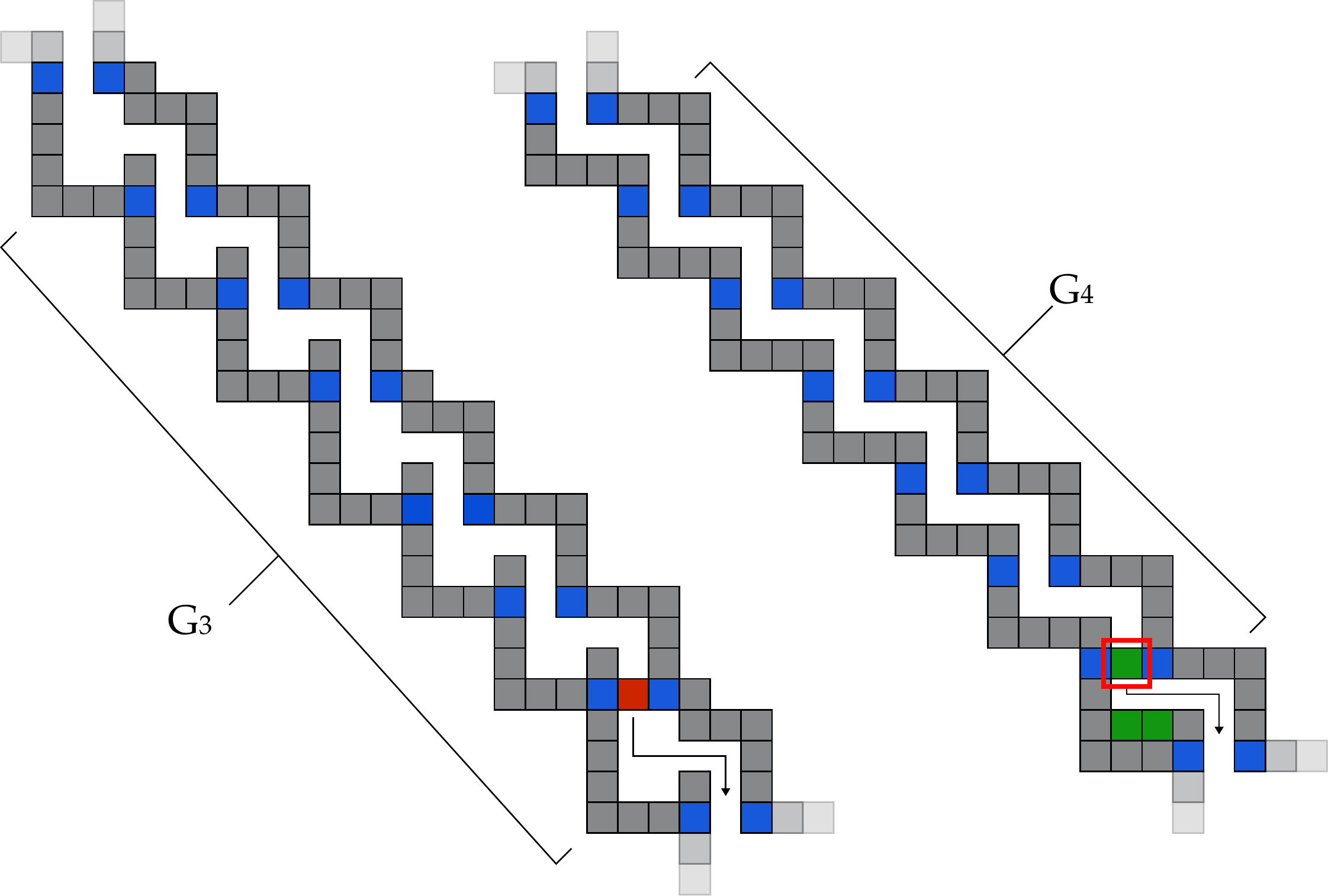}
    \text{(c)}
  \end{minipage}
  \centering
  \caption{During the second phase, unconfined literal tiles are forced inside the Notch gadget of the fourth chain of their corresponding clause chain. Unsatisfied clause chains export one too many literal tiles, as shown in (c).}
  \label{fig:2phase}
\end{figure}

Afterwards, the relocation and any unconfine literal tiles will reside inside the third and fourth chain, respectively, of their corresponding section/chain, as depicted in Figure \ref{fig:2phase}.
The last gadget of the fourth chain of every clause chain consists of a Notch gadget, whose function in this case is to provide two open spaces for the unconfined literal tiles from the clause chains.
Moving the relocation tile through the third chain of the assignment section, each remaining literal tile will reach the last Notch gadget of their corresponding clause chain and fit in one of the two open spaces made by the Notch gadget, provided that the clause chain was satisfied.
%If a clause chain was unsatisfied, then as the relocation tile traverses through the last Assignment gadget of the third chain of the assignment section the third literal tile will not fit within the two open spaces provided by the Notch gadget, keeping it unconfined.
If a clause chain was unsatisfied, then the third literal like will not fit within the two open spaces provided by the Notch gadget, leaving it free to continue through the board.
At this point, the relocation tile would have reached the fourth chain of the assignment section, which is composed entirely of Notch gadgets and whose length is equal to that of the length of the formula section plus the length of the validation section.
%Traversing the relocation tile through the fourth chain will simultaneously move any unconfined literal tile inexorably towards the validation section and into location $b$, given that the fourth chain is sufficiently large enough to have such an effect.
Traversing the tile through the fourth chain will move any free literal tiles towards the validation section, blocking location $b$.
Therefore, if there exist at least one unsatisfied clause chain, there exists at least one unconfined literal tile inexorably pushed to occupy location $b$ whilst the relocation tile traverses through the architecture of the fourth chain of the assignment section.
If there are no unsatisfied clause chains, it follows that there will be a direct path from the relocation tile to location $b$ on the board allowing its relocation there.

\subsection{Theorem}
\begin{theorem}\label{thm:monotone_relocation}
The relocation problem under the step transformation is NP-Complete on a monotone board when limited to two directions.
\end{theorem}
\begin{proof}
Membership in NP is described in Section \ref{sec:np}.
To show NP-Completeness, a reduction from 3SAT is detailed such to show relocation in monotone boards solves 3SAT.
For a given 3SAT instance, a monotone board $B = (O, W)$ is designed such that the elements of the 3SAT instance are represented by corresponding elements of the monotone board.
The elements of a given 3SAT instance consist of the set of clauses along with the literals that make up each individual clause.
The literals of each clause are represented by individual tiles, which inhabit their pertaining clause chain.
As stated above, the literals can be assigned some truth value by either confining the associated tile within its pertaining clause chain (e.g, assigning $true$) or keeping it unconfined (e.g, assigning $false$).
The reduction can be understood as a two phase process:

\textbf{Phase one.}
Starting with an initial board $B = (O, W)$ described in the previous section, the first phase consists of making variable truth assignments in ascending order for the set of boolean variables $X = \{x_1, x_2, \ldots, x_N\}$.
Each variable can be given some truth value by performing either step sequence illustrated in Figure \ref{fig:trueFalse}.
Simultaneously, each individual literal tile in the clause chains will reach a corresponding Positive or Negative gadget exactly when the variable it equals is assigned some truth value.
When assigning a variable to $true$, the positive literals that equal that variable will become confined, whereas the negative literals that equal that variable will remain unconfined.
Likewise, assigning a variable to $false$ keeps the positive literals unconfined and the negative ones confined.
During this phase, the design of the validation section along with the inhabitant validation tiles restrict the available step sequences to that of the two previously mentioned.
Given that the Goal gadget resides within the validation section, along with location $b$ of the board, the validation tiles risk occupying location $b$ unless either of the step sequences are performed during this phase.
The restrictive nature of the validation section ensures that phase one of the reduction consists only of valid variable truth assignments.

\textbf{Phase two.}
Afterwards, every clause chain assumes some satisfiability status by the quantity of confined literal tiles residing within them.
During this phase, the number of unconfined literal tiles of each clause chain are `counted' by the attempt to occupy two open spaces inside their corresponding clause chains with these literal tiles.
A satisfied clause chain will confine at least one literal tile and leave at most two literal tiles unconfined.
These two unconfined literal tiles are able to occupy the two open spaces, essentially removing them from the path to location $b$ from location $a$.
On the other hand, an unsatisfied clause yields one too many unconfined literal tiles for the two corresponding open spaces, causing one of the literal tile to remain within the path to location $b$ from location $a$.

Following this two phase process, the tile initialized at location $a$ is relocatable to location $b$ on the board if every clause chain yielded at most two unconfined literal tiles after the first phase.
During the second phase, these two unconfined literal tiles can be removed from the path from location $a$ to location $b$, allowing access for the relocation tile to the targeted location on the board.
Similarly, relocation becomes impossible when a clause chain produces three unconfined literal tiles after phase one since the attempt to relocate after the second phase will inexorably occupy location $b$ with one of the unconfined literal tiles.
It follows that the relocation of the relocation tile initialized at location $a$ is possible if all clause chains were satisfied, which is possible only with a satisfying truth assignment to the set of variables.
Moreover, if the boolean formula is not satisfiable, then some clause chains will always yield three unconfined literal tiles. The forward-progressing nature of the construction, along with the inability to store tiles in improper notches, means that the relocation tile can never be relocated to its goal position (because it is blocked).
Therefore, for a given 3SAT instance relocation is possible if and only if the boolean formula is satisfiable.
\end{proof}

\begin{theorem}
The relocation problem under the step transformation is NP-Complete on a monotone board when limited to three directions.
\end{theorem}
\begin{proof}
The inclusion of the third direction $west$ does not change the phases of the reduction though it allows for cycling between different board configurations reachable by $east$ and $west$ movements.
During phase one, the validation section is used to restrict the step sequences available when assigning truth values to variables where the inclusion of the $west$ direction only adds the cycling between the choices of the step sequences (assigning $true$ or assigning $false$).
Thus, adding the $west$ direction does not change the way variable truth assignments are performed.
Moreover, the manner in which tiles traverse through the board remains functional since the movement in the $west$ direction simply reverts the board back to the configuration that was previously visited when a $east$ direction was performed.
Thus, the phases of the reduction persist even with the third direction.
\end{proof}

%% file: reconfiguration.tex
\section{Reconfiguration with Limited Directions}
We now will take a look at the problem of reconfiguration on a connected board when limited to two orthogonal directions. We show that with these constraints the problem is NP-Complete under the step transformation. Without loss of generality we will be limiting the directions to south and east. %We will present individual gadgets which can be connected in a way that does not effect their functionality.
In this section, we prove hardness with a reduction from 3SAT. %Assume we are given a 3SAT problem consisting of variables $x_1, x_2, x_3, ... , x_n$, separated into $m$ clauses of the form $A \lor B \lor C$ where each $A,B,C$ is of the form $x_i$ or $\neg x_i$. We create an instance of the reconfiguration problem that is solvable if and only if the respective 3SAT problem is solvable.
The following subsections describe the gadgets used when constructing a tilt configuration from a 3SAT formula with $n$ variables and $m$ clauses.

\begin{figure}[tb]
  \begin{minipage}[t]{1.\columnwidth}
    \centering
    \includegraphics[width=1.\textwidth]{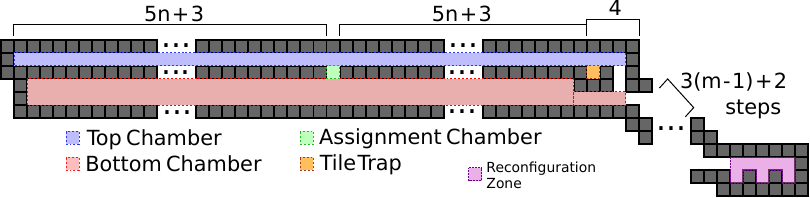}
    \\ \text{(a)}
  \end{minipage}
  \begin{minipage}[t]{1.\columnwidth}
    \centering
    \includegraphics[width=1.\textwidth]{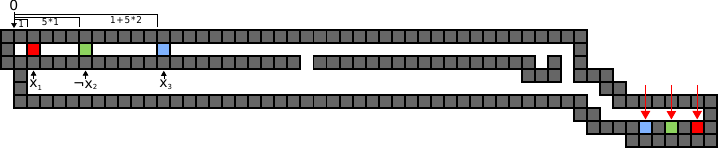}
    \\ \text{(b)}
  \end{minipage}
  \caption{(a)3 SAT clause gadget, where $n$ is the total number of unique variables, and $m$ indicates that this clause is the $m$th clause in the formula. (b) Example of variable placement for clause $(x_1 \lor \neg x_2 \lor x_3)$ in a 5 variable formula. Goal locations indicated by red arrows.}
  \label{fig:reconf_clause}
  \label{fig:reconf_variable_placement}
\end{figure}

\subsection{Gadget Construction}\label{subsec:recon_gadget_const}
In this subsection, we describe how to construct an instance of the reconfiguration problem from an instance of the 3SAT problem. We discuss this construction in terms of gadgets.

\paragraph{Clause Gadget}
Each clause of the 3SAT formula will be represented by a clause gadget (depicted in Figure~\ref{fig:reconf_clause}). This gadget consists of a top chamber in which variable tiles will be placed and a 1-wide gap that connects that chamber to a second bottom chamber.  There will be a staircase that connects the lower chamber to a final reconfiguration zone, which contains the goal locations for the three variable tiles in the clause.  The number of steps of the staircase will be a function of the current clause; so, the $m^{\text{th}}$ clause in the formula will have a staircase with $3(m-1) + 2$ steps.

For every variable $x_i$, in every clause that it appears, we will place a corresponding variable tile in the top chamber of its respective clause gadget at position $5(i-1) + 1$ for every positive variable and $5(i-1)$ for every negated variable. The goal locations of the tiles will be one in each of the notches. Without loss of generality, the goal locations will be in the reverse order from west to east that they start at in the top chamber. Figure \ref{fig:reconf_variable_placement} shows an example tile placement for a clause gadget.

\begin{figure}[tb]
  \begin{minipage}[t]{.49\columnwidth}
    \centering
    \includegraphics[width=1.\textwidth]{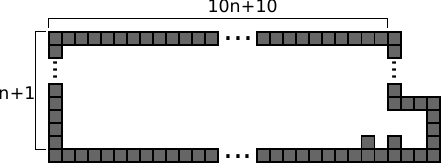}
    \\ \text{(a)}
  \end{minipage}
  \begin{minipage}[t]{.49\columnwidth}
    \centering
    \includegraphics[width=1.\textwidth]{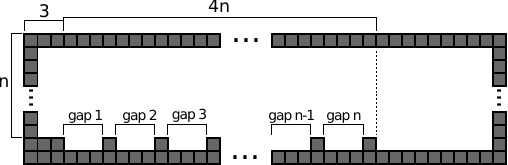}
    \\ \text{(b)}
  \end{minipage}
  \caption{(a) South Limiter: Limits the amount of south steps made before all variables have been assigned. (b) South Forcer: Forces the user to make south steps at specific times}
  \label{fig:reconf_down_limiter}
  \label{fig:reconf_down_forcer}
\end{figure}

\begin{figure}[tb]
  \begin{minipage}[t]{1.\columnwidth}
    \centering
    \includegraphics[width=.9\textwidth]{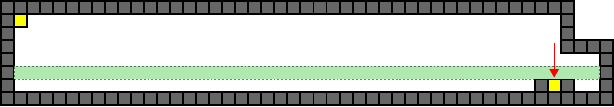}
    \\ \text{(a)}
  \end{minipage}
  \begin{minipage}[t]{1.\columnwidth}
    \centering
    \includegraphics[width=.9\textwidth]{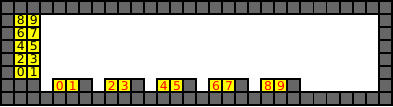}
    \\ \text{(b)}
  \end{minipage}
  \caption{(a) Example of south limiter tile placement in a 5 variable formula. Goal location indicated by red arrow. Post assignment zone highlighted in green. (b) Example of south forcer tile placement in a 5 variable formula. Goal location indicated by red labels.}
  \label{fig:down_limiter_tile_placement}
  \label{fig:down_forcer_tile_placement}
\end{figure}

%\begin{figure*}
%\centering
%\begin{minipage}{.5\textwidth}
%  \centering
%  \captionsetup{width=.75\linewidth}
%  \includegraphics[width=.9\linewidth]{images/reconf/limiter1labeled.pdf}
%  \captionof{figure}{South Limiter: Limits the amount of south steps made before all variables have been assigned}
%  \label{fig:reconf_down_limiter}
%\end{minipage}%
%\begin{minipage}{.5\textwidth}
%  \centering
%  \captionsetup{width=.75\linewidth}
%  \includegraphics[width=.9\linewidth]{images/reconf/limiter2labeled.pdf}
%  \captionof{figure}{South Forcer: Forces the user to make south steps at specific times}
%  \label{fig:reconf_down_forcer}
%\end{minipage}
%\end{figure*}

\paragraph{South Limiter Gadget}
Every instance of the reconfiguration problem obtained through this reduction will contain a single \emph{south limiter gadget} (show in Figure~\ref{fig:reconf_down_limiter}). The purpose of this gadget is to limit the number of \emph{south} steps that can be made in the reconfiguration step sequence. The height and width of this gadget will scale linearly to the number of variables in the 3SAT formula.

In the South Limiter gadget we  place a tile at the northwest corner, and the goal location is the notch at the other end of the gadget. Figure \ref{fig:down_limiter_tile_placement} depicts tile and goal placement for this gadget.

\paragraph{South Forcer Gadget}
Each instance will also have one \emph{south forcer gadget} (shown in Figure~\ref{fig:reconf_down_forcer}). The purpose of this gadget is to enforce the proper timing of any \emph{south} step made in the reconfiguration step sequence. The height and width of this gadget also scales linearly to the number of variables.

In the South Forcer gadget we  place 2 tiles in a row for every distinct variable in the 3SAT formula on the west side of the gadget. The goal location of row $n$ is the eastmost side of the $n^{th}$ gap \ref{fig:reconf_down_limiter}, in the same position relative to each other that they started in. Figure \ref{fig:down_forcer_tile_placement} depicts tile and goal placement for this gadget.

\subsection{Gadget Functionality}\label{subsec:recon_gadget_func}
Here, we discuss the relationship between our tilt reconfiguration and the 3SAT variable assignment, as well as the functionality of each of the gadgets presented in the previous subsection.

\paragraph{Variable Assignment}
This reduction works by utilizing the uniform global signals to assign all variables $x_i$ a truth value simultaneously. To start the process of reconfiguration we will begin stepping east, uniformly shifting all variable tiles along the top chamber of the clause gadgets. When the eastmost variable tile is located above the assignment chamber, the choice of assigning that variable true can be made by performing step sequence $s_t = \langle S,E \rangle$, or false with the step sequence $s_f = \langle E,S \rangle$. A variable tile evaluates to true if it enters the assignment chamber.
%Since for every clause, every variable tile representing a certain variable share the same $x$ coordinate, all of these variable tiles will receive the same assignment.
Since all variable tiles $t_i$ representing a particular variable $x_i$ share the same x-coordinate, all of these tiles will receive the same assignment.

\paragraph{South Forcer/Limiter Gadgets}
The reconfiguration requirement of the \emph{south forcer gadget} ensures that every variable must receive an assignment. The only way to place each tile-pair in their respective goal positions is to ``assign'' each of the $n$ variables a value of either true of false.
The \emph{south limiter gadget} ensures that only \emph{n} assignments can be made, since doing more would position the gadget's tile along the bottom edge of the gadget with no way of reaching its goal position.
The combination of these gadgets prevents the possibility of assigning both a positive and negated variable tile, i.e. $t_i$ and $\neg t_i$, values of ``true''.
%The \emph{south limiter gadget} ensures that only \emph{n} assignments can be made, since doing more would leave that gadget's tile in an unreconfigurable spot.
Once every variable tile has been assigned, the \emph{south limiter gadget's} tile will be in the \emph{post assignment zone} (see Figure~\ref{fig:down_limiter_tile_placement}). In order to achieve the configuration goal inside that gadget, only east steps can be made until the tile is directly above its goal location. With global signals this causes all variables tiles to be pushed maximally to the east side of the top chamber of their respective clause gadget. Once the \emph{south limiter's} tile is above its goal location, a south step must be performed, since we are limited to only south and east, and an east step would make the south limiter's tile pass its goal position.

\begin{figure}[tb]
    \centering
        \includegraphics[width=1.\columnwidth]{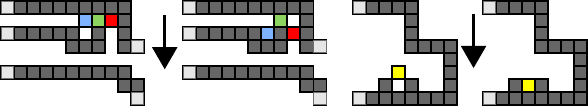}
        \caption{State of a clause gadget in which no variable tiles evaluated to a literal true before and after the forced south tilt, and the respective states of the \emph{south limiter gadget}}
    \label{fig:reconf_trapped_variable_tiles}
\end{figure}

\paragraph{Clause Verification}
A clause gadget is in its goal configuration when each of its variable tiles are in their goal locations at the bottom right of the gadget.
A clause gadget will be unable to reach its goal configuration if no variable tiles make it through the assignment chamber (this is the equivalent to every literal of a 3SAT clause evaluating to false).
If there are three variable tiles still in the top chamber of a clause gadget, a south step will place the west-most variable tile into the \emph{tile trap}. This leaves the board unreconfigurable (see Figure~\ref{fig:reconf_trapped_variable_tiles}).
If at least 1 variable tile is passed into the assignment chamber, then the clause gadget will always be reconfigurable.
From there the tiles can be moved to their respective staircase, where the varied number of steps allow for each tile to moved to its goal location individually.

\subsection{Formal Proof}
%In this subsection, we provide the formal proof for this problem.

\begin{theorem} \label{thm:2dir_reconfig}
The reconfiguration problem under the step transformation is NP-Complete on a connected board when limited to two directions.
\end{theorem}

\begin{proof}
%Assume we are given a 3SAT problem consisting of variables $x_1, x_2, x_3, ... , x_n$, separated into $m$ clauses of the form $A \lor B \lor C$ where each $A,B,C$ is of the form $x_i$ or $\neg x_i$. We create an instance of the reconfiguration problem that is solvable if and only if the respective 3SAT problem is solvable.
To show hardness for the reconfiguration problem under the step transformation, we reduce from 3SAT.
Given a 3SAT formula consisting of $n$ variables and $m$ clauses, we construct a tilt configuration using the gadgets described in Section~\ref{subsec:recon_gadget_const}. We create a clause gadget for every clause in the 3SAT formula, as well as a single South Forcer Gadget and a single South Limiter Gadget. To make this a connected board these gadgets can be attached at their northwest corner without affecting their functionality.

First, we argue that the instance of the reconfiguration problem obtained through this reduction is solvable if the respective 3SAT formula is solvable. This follows from the variable assignment paragraph in Section~\ref{subsec:recon_gadget_func}. Given a certificate containing the truth values that satisfy the 3SAT formula, start the reconfiguration by stepping east. When variable tile $t_i$ is up for assignment, i.e., the positive variable tile representing $x_i$ is above the assignment chamber, a $\langle S,E \rangle$ step sequence can be input if the variable $x_i$ is being set true, and a $\langle E,S \rangle$ step sequence if the variable $x_i$ is to be assigned false. %If the certificate contains a valid solution to the 3SAT formula, then at least one tile in every clause will evaluate to true, making the board reconfigurable to its goal configuration.
Since the certificate contains a valid solution to the 3SAT formula, then at least one tile in every clause will evaluate to true (following the functionality section), meaning all clause tiles can be placed in their goal locations. Furthermore, the tiles in the south limiter gadget and the south forcer gadget will reach their goal positions via the standard variable assignment.

Next, we argue that the respective 3SAT problem is solvable if the obtained instance of the reconfiguration problem has a solution. This direction relies on the functionality of the South Limiter and South Forcer gadgets, as well as the clause verification discussed in Section~\ref{subsec:recon_gadget_func}.
We define $d_a$ as the distance between the westmost variable tile ($x_1$ or $\neg x_1$ if it exists) and the assignment chamber. We  observe the distance from the tile in the \emph{south limiter gadget} to its goal location is greater than $d_a$. Only \emph{n} south steps can be input before the south limiter gadget's tile is above its goal location. Every variable tile assignment requires a south step, and once the \emph{south limiters gadget's} tile is above its goal location, all variable tiles are no longer assignable. Thus, we can only give an assignment to \emph{n} variable tiles. The issue remains that we can assign both a positive and negated variable tile, i.e. $t_i$ and $\neg t_i$, and skip the assignment of some other variable tile $t_j$. The \emph{south forcer gadget} ensures this can not happen by requiring south steps to be made at certain times in order to satisfy the reconfiguration requirements. This restriction ensures that each variable must receive a a true or false assignment.

Thus, the generated instance of the reconfiguration problem is solvable if and only if the given 3SAT formula is solvable.
Along with Thm. \ref{thm:NP}, it follows that the %reconfiguration 
problem %(under unit movements) 
is NP-Complete when limited to two directions.
\end{proof}

%% file: conclusion.tex
\section{Conclusion and Future Work}

In this paper we investigated the Occupancy, Relocation, and Reconfiguration problems in an extremely limited variation of the tilt model. We discovered that even with these limitations, the problems of Relocation and Reconfiguration were still NP-Complete under the step transformation. Future work may focus on pushing the limitations even further; possibly considering relocation and/or reconfiguration of a rectangular board.